\documentclass[prb,twocolumn,showpacs,preprintnumbers,amsmath,amssymb]{revtex4}

\usepackage{graphicx}
\usepackage{dcolumn}
\usepackage{bm}

\begin{document}

\title{Suppression of low-energy Andreev states by a supercurrent in YBa$_2$Cu$_3$O$_{7-\delta}$}

\author{J. Ngai}
 \affiliation{Department of Physics, University of Toronto, 60 St. George St., Toronto, ON  M5S1A7, Canada}
\author{P. Morales}
 \affiliation{Department of Physics, University of Toronto, 60 St. George St., Toronto, ON  M5S1A7, Canada}
\author{J.Y.T. Wei}%
 \affiliation{Department of Physics, University of Toronto, 60 St. George St., Toronto, ON  M5S1A7, Canada}
\date{\today}

\begin{abstract}
We report a coherence-length scale phenomenon related to how the
high-$T_c$ order parameter (OP) evolves under a
\emph{directly-applied} supercurrent. Scanning tunneling
spectroscopy was performed on current-carrying
YBa$_2$Cu$_3$O$_{7-\delta}$ thin-film strips at 4.2K. At current
levels well below the theoretical depairing limit, the low-energy
Andreev states are suppressed by the supercurrent, while the
gap-like structures remain unchanged. We rule out the likelihood
of various extrinsic effects, and propose instead a model based on
phase fluctuations in the \emph{d}-wave BTK formalism to explain
the suppression. Our results suggest that a supercurrent could
weaken the \emph{local} phase coherence while preserving the
pairing amplitude. Other possible scenarios which may cause the
observed phenomenon are also discussed.
\end{abstract}


\maketitle

\section{INTRODUCTION}

High critical-temperature ($\emph{T}_c$) cuprate superconductors
are distinguished from conventional \emph{s}-wave superconductors
by a predominantly \emph{d}-wave order parameter
(OP)\cite{Tsuei,VanHarlingen}.  The \emph{d}-wave OP symmetry has
led to a wealth of unconventional experimental phenomena. One such
phenomenon is the formation of low-energy Andreev states bound on
non principal-axis surfaces of cuprate superconductors
\cite{Hu,TanakaKashiwaya,ABSrevu}. When probed by quasiparticle
tunneling, these Andreev states are manifested as zero-bias peaks
(ZBCP) in the conductance spectra of high-impedance tunnel
junctions. These Andreev states form when time-reversed
quasiparticles interfere constructively via the phase sign change
of the \emph{d}-wave OP about its nodal ($k_x$=$\pm$$k_y$)
axes\cite{Hu,TanakaKashiwaya}. For \emph{s}-wave superconductors,
tunneling involves quasiparticle transmission, and is thus only
sensitive to the amplitude of the OP. Tunneling into \emph{d}-wave
superconductors involves both quasiparticle transmission and
Andreev reflection, and is thus sensitive to both the phase and
amplitude of the pair
wavefunction\cite{Hu,TanakaKashiwaya,WeiPRL}.

Recent interest has been focused on how the \emph{d}-wave Andreev
bound states, quasiparticle density of states, and pairing
symmetry itself could evolve under an applied
supercurrent\cite{Doppler,Khavkine,Zapotocky,Kabanov}. In this
paper, we present a scanning tunneling spectroscopy (STS) study of
YBa$_2$Cu$_3$O$_{7-\delta}$ (YBCO) thin-film strips carrying
directly-applied supercurrents at 4.2K. Our study uses the
supercurrent to perturb the superconducting OP
\emph{electrodynamically}, rather than \emph{thermodynamically} as
was previously done in magnetic field studies\cite{Aprili,Krupke}.
High-$\emph{T}_c$ cuprates have characteristically short coherence
lengths, which can be $\sim$nm in the basal plane.  By virtue of
its nanoscale junction size, STS can thus provide coherence-length
scale information about the high-$\emph{T}_c$ OP when driven by an
applied supercurrent. In our experiment, we observed systematic
suppression of the \emph{phase}-sensitive spectral features by the
supercurrent, while the \emph{amplitude}-dependent features remain
largely unchanged. We discuss and rule out the likelihood of
various extrinsic effects, and propose instead a model based on
current-driven OP phase fluctuations to describe the phenomenon
observed.

\section{EXPERIMENTAL SETUP AND RESULTS}

\begin{figure}[t]
\includegraphics[width=3.3in] {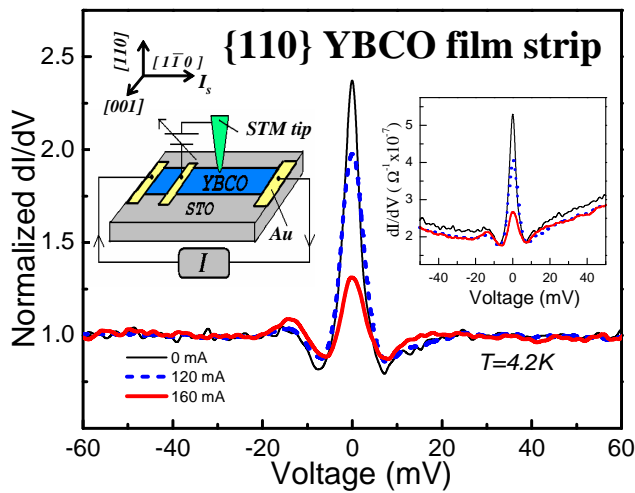}
\caption{\label{PRL_fig1}(Color online) Normalized \emph{dI/dV}
spectra obtained on a \{110\} YBCO film strip carrying a
supercurrent \emph{I}$_s$ applied along $[$1$\overline{1}$0$]$ at
4.2K. Right inset displays the unnormalized data. Left inset
displays our experimental setup showing an STM tip over a YBCO
thin-film strip grown on an STO substrate with the orientation as
indicated. The tip is biased relative to the sample by a gold pad on
the sample surface. Spectroscopy measurements are made while a
simultaneous current is passed through the film strip.}
\end{figure}

Our experimental setup is illustrated in the left inset of Figure
\ref{PRL_fig1}, showing a Pt-Ir tip positioned over a YBCO
thin-film strip in an STS geometry, with a Au contact providing
the bias voltage \emph{V} between tip and sample. STS is performed
in the usual manner, by measuring the tunneling current
\emph{I}$_t$ with the scanning and feedback temporarily suspended,
but while a current \emph{I}$_s$ is simultaneously applied through
the superconducting film at 4.2K.  In this STS geometry, the high
junction impedance ($\sim$0.1G$\Omega$) ensures that \emph{I}$_t$
($\sim$1nA) is decoupled from \emph{I}$_s$ ($\sim$100mA). The
\emph{I}$_s$ is supplied from a floating source below, via Au
contacts on the ends of the film strip.  A special circuitry is
used to synchronize \emph{I}$_s$ with the STS feedback and data
acquisition \cite{NgaiAPL}.  Short duty-cycle \emph{I}$_s$ pulses
$\approx$200$\mu$s in width are used to prevent sample Joule
heating, which could occur at high current levels.  DC
measurements are also made at lower current levels for comparison,
in order to rule out transient electronic effects. For our
experiment, this versatile STS technique enabled very low-noise
($<$5pA) and highly reproducible tunneling $I$-$V$ measurements at
nanometer length scales.

The YBCO thin-film strips used in our experiment were grown by
pulsed laser-ablated deposition on SrTiO$_{3}$ (STO) substrates.
Both \{110\} and \{001\} oriented films were made, the former
using a templating technique.  The films were typically 50nm thick
and in a 1mm x 3mm strip-line geometry, aligned along $<$100$>$
and $[$1$\overline{1}$0$]$ respectively for the \{001\} and
\{110\} films. The film strips showed well-defined $\emph{T}_c$'s
($\approx$87K) and sizable transport critical-current densities at
4.2K ($\emph{J}_c$$\approx$5$\times$$10^6$A/cm$^2$, defined by the
appearance of 1$\mu$V/mm along the strip). Epitaxiality of the
films was confirmed by x-ray diffraction and rocking-curve
analysis. Film surface quality was determined directly from
topography measurements using our STS setup.  Shown in the inset
of Figure \ref{PRL_fig4} is a topographic image taken on a \{001\}
film, indicating smooth terraces with large \emph{c}-axis faces
$\sim$30nm in size and \emph{ab}-plane edges $\sim$3nm in height.

Figure \ref{PRL_fig1} shows tunneling conductance data taken at 4.2K
on a \{110\} film strip at various current levels applied along
$[$1$\overline{1}$0$]$.  The $dI/dV$ spectra, which were numerically
differentiated from the $I$-$V$ data, are plotted in the right
inset, and the normalized $dI/dV$ spectra are plotted in the main
panel. These spectra show a well-developed ZBCP structure which is
suppressed in height but only slightly changed in width by the
applied current. Such ZBCP structures have been commonly observed in
YBCO, and could be identified as the low-energy Andreev states based
on a predominant \emph{d}-wave OP \cite{Hu,TanakaKashiwaya,WeiPRL}.
Two features of the ZBCP spectral evolution versus current are
noteworthy. First, the ZBCP shows no splitting, indicating
time-reversal invariance of the \emph{d}-wave OP up to 160mA (the
maximum applied current) in our near-optimally doped YBCO samples at
4.2K. Second, the area under \emph{dI/dV} is not conserved,
indicating negligible thermal broadening, with a clear loss of
Andreev states. The non-splitting of the ZBCP we observed is
quantitatively consistent with a recent theoretical calculation
\cite{Doppler}, which shows no Doppler shift \cite{Aprili,Krupke} of
the Andreev bound states in this geometry until a higher
supercurrent density.  However, the clear loss of Andreev states we
observed was not anticipated by this mean-field calculation.

\section{ANALYSIS}

We have considered various extrinsic effects in analyzing the
phenomenon we observed. One possible extrinsic effect is the Joule
heating from the sizable current applied. However, our pulsed
current technique, along with the absence of significant ZBCP
broadening, largely rule out this possibility. A second possible
extrinsic effect involves current-generated vortices that could
introduce Aharonov-Bohm phases in the Andreev interference
process\cite{Asano}. We note that our current levels are well
below the theoretical depairing limit\cite{Doppler}, and also
below any sharp voltage onsets associated with vortex depinning,
although our present experimental resolution cannot yet rule out
the presence of such vortices. Nevertheless, the theoretical
calculation in Ref.14 has predicted a ZBCP broadening which we do
not observe, thus rendering this vortex scenario unlikely. A third
possible extrinsic effect involves excess low-energy
quasiparticles due to the applied nodal current\cite{Yip}. In this
scenario, the excess quasiparticles could effectively raise the
quasiparticle temperature, again creating spectral broadening
which we do not observe. In summary, all three extrinsic effects
described above would be expected to yield a conservation of the
low-energy Andreev states, and are therefore unlikely to explain
our observations.

Having ruled out the likelihood of these various extrinsic
scenarios, we propose instead an explanation based on
current-driven OP phase fluctuations to describe the observed
phenomenon. High-temperature superconducting cuprates are
distinguished by their low carrier densities and short coherence
lengths\cite{EmeryKivelson}. These properties imply an inherently
low superfluid stiffness, making the order parameter susceptible
to fluctuations in its phase \cite{Merchant}. Motivated by this
possibility, we incorporate phase fluctuations into the
generalized Blonder-Tinkham-Klapwijk (BTK)\cite{Blonder}
formalism, which has been theoretically
established\cite{TanakaKashiwaya} for \emph{d}-wave
superconductors, and experimentally verified for
YBCO\cite{WeiPRL}. The generalized BTK expression for tunneling
conductance \emph{G}$_{ns}$ is given
by\cite{Hu,TanakaKashiwaya,WeiPRL}:

\begin{eqnarray}
\label{BTK}
    \frac{G_{ns}}{G_{nn}}=\int^{+\frac{\pi}{2}}_{-\frac{\pi}{2}}d\theta\int^{+\infty}_{-\infty} dE[1+|
    a|^2-|b|^2] \frac{\partial\textsl{f}(E-eV)}{\partial V}
\end{eqnarray}
where $a$ and $b$ are the Andreev-reflection and normal-reflection
coefficients, $\textsl{f}$ is the Fermi-Dirac function,
$\emph{G}_{nn}$ is the normal-state junction conductance, \emph{E}
the quasiparticle energy, and $\theta$ the polar angle in
\emph{k}-space. To model the effects of OP phase fluctuations, we
modify the BTK kernel from Ref.4 by adding a phase factor
e$^{i\varphi}$ to the phase-interference term in its denominator
(see Eqn.\ref{intBTK} below) then integrating it over the domain of
phase $\varphi$ weighted by a Gaussian of width $\gamma$:
\begin{eqnarray}
\label{intBTK}
    1+|a|^2-|b|^2=\int^{+\pi}_{-\pi}d\varphi\frac{1}{\sqrt{2\pi\gamma}} e^{-\varphi^{2}/\gamma^{2}}\times\nonumber\\
    \frac{16(1+|\Gamma_+|^2)\cos^4\theta+4Z^2(1-|\Gamma_+\Gamma_-|^2)\cos^2\theta}{|4\cos^2\theta+Z^2[1-\Gamma_+\Gamma_-e^{i\phi_--i\phi_++i\varphi}]|^2}
\end{eqnarray}

Note that the amplitude factors $\Gamma_\pm =
(E/|\Delta_\pm|)-\sqrt{(E/{|\Delta_\pm|})^2-1}$ contain a
\emph{d}-wave gap function with $\Delta_0$ as gap maximum. The
phase factors $e^{i\phi_\pm}={\Delta_\pm}/{|\Delta_\pm|}$
represent the sign of the pair potential
$\Delta_\pm=\Delta(\theta_\pm)$ which is experienced by an
Andreev-reflected quasielectron (or quasihole) propagating at an
angle $\theta_+$ (or $\theta_- = \pi - \theta_+ )$ relative to the
junction normal (see left inset in Fig.3). Through the extra phase
introduced by $e^{i\varphi}$, the weighted integral over $\varphi$
essentially smears out the \emph{relative} phase between
\emph{consecutive} Andreev-reflected quasiparticles. This phase
smearing would cause incoming and outgoing quasiparticles to
interfere less constructively, thereby leading to a loss of the
low-energy Andreev states. Our phenomenological approach can be
compared with the more rigorous treatment of Ref.19 which uses a
2D X-Y model based on vortex fluctuations \cite{FranzMillis}. Our
model makes no assumptions about which physical mechanism is
actually driving the phase fluctuations, while also going beyond
the low-impedance (\emph{Z}$\ll$1) and principle-axis (\{100\})
junction cases\cite{ChoiCampbell}.

\begin{figure}[t]
\includegraphics[width=3.3in] {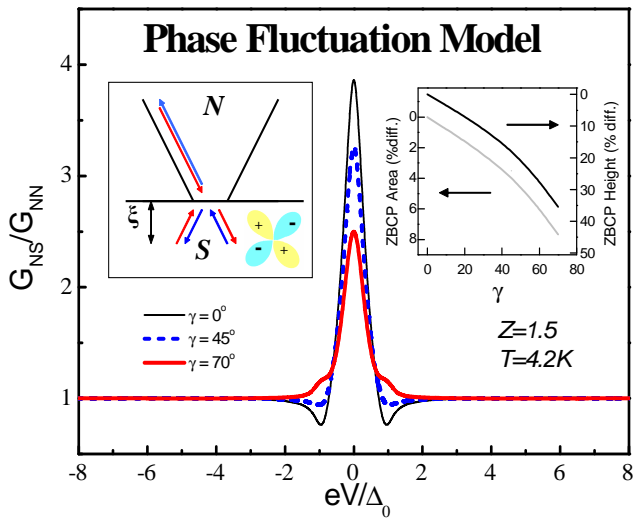}
\caption{\label{PRL_fig2}(Color online) Phase fluctuation model
showing the evolution of the ZBCP on a \{110\} junction as a
function of the phase smearing parameter $\gamma$. Left inset
illustrates the Andreev interference process based on the
\emph{d}-wave phase sign change. Right inset shows the ZBCP height
and spectral area as a function of $\gamma$.}
\end{figure}

Our spectral simulations for a \{110\} junction are plotted in
Figure \ref{PRL_fig2}, as normalized conductance
\emph{G}$_{ns}$/\emph{G}$_{nn}$. A large barrier parameter
\emph{Z}=1.5 was used to represent our high-impedance junctions,
along with the experimental temperature \emph{T}=4.2K and a typical
gap maximum $\Delta_0$=15meV.  A clear trend of ZBCP suppression is
seen versus the phase-smearing parameter $\gamma$, showing a drop in
the peak height and a loss of spectral area. Detailed evolutions of
height and area versus $\gamma$ are plotted in the right inset of
Fig.\ref{PRL_fig2}.  Comparing our data in Fig.\ref{PRL_fig1} with
the simulations in Figure \ref{PRL_fig2}, it is clear that the ZBCP
suppression could be qualitatively attributed to the
\emph{dephasing} effects of current-driven OP phase fluctuations.
Physically speaking, consecutive Andreev-reflected quasiparticles
which are bound within a coherence length $\xi$ from the specular
\{110\} surface (see left inset of Fig.2), would experience an
effective smearing in their relative phase as a result of the phase
fluctuations, thus interfering less constructively. This phase
smearing leads directly to a loss in the low-energy Andreev states,
resulting in the ZBCP suppression seen in the data.

\begin{figure}[t]
\includegraphics[width=3.3in] {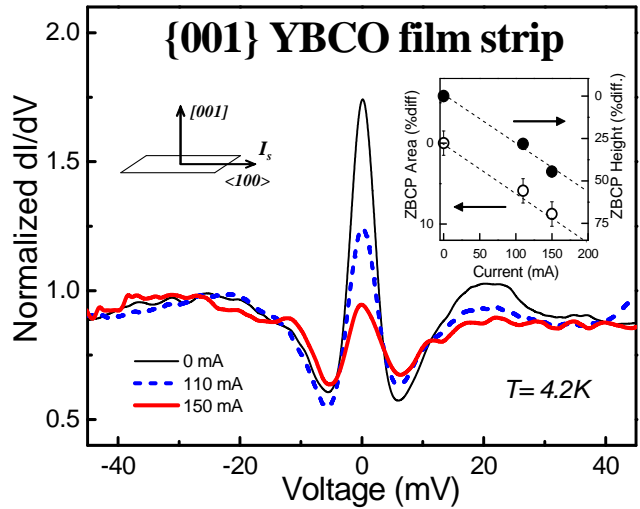}
\caption{\label{PRL_fig3} (Color online) Normalized \emph{dI/dV}
spectra obtained on a \{001\} YBCO thin-film strip carrying a
supercurrent \emph{I}$_s$ applied predominantly along $<$100$>$ at
4.2K. Right inset shows the evolution of the ZBCP height and
spectral area as a function of the applied current.}
\end{figure}

\begin{figure}[t]
\includegraphics[width=3.3in] {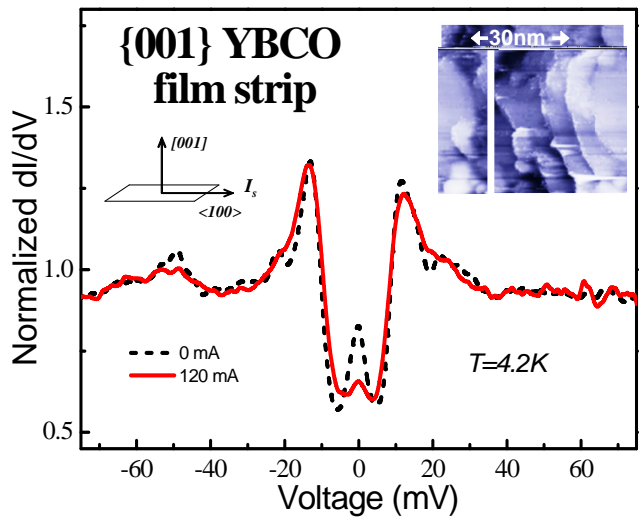}
\caption{\label{PRL_fig4} (Color online) Normalized \emph{dI/dV}
spectra obtained on a \{001\} YBCO thin-film strip carrying a
supercurrent \emph{I}$_s$ applied predominantly along $<$100$>$ at
4.2K. The inset shows film topography.}
\end{figure}

\section{DISCUSSION}

It is important to consider possible physical mechanisms by which
the supercurrent is driving the local phase fluctuations. One
scenario involves the possibility of spatial doping
inhomogeneities in the superconducting cuprates\cite{LangDavis}.
In the presence of current flow, local areas of strong
superconductivity could become decoupled from each other, thus
permitting local phase fluctuations. Another scenario may involve
the proliferation of thermally-generated vortex-antivortex pairs
in the Kosterlitz-Thouless picture \cite{Minnhagen,QED3}, but with
current as an additional driving force \cite{Beasley,Epstein}. In
principle, the \emph{static} phase gradient carried by the
supercurrent could also couple \emph{dynamically} with other types
of phase-ordered states, such as spin/charge-density waves
\cite{DemlerZhang,deLozanne}, staggered flux-like states
\cite{Affleck,Varma,DDW} or collective modes
\cite{KimWen,Devillard}, to produce enhanced phase fluctuations.
In general, any of the above scenarios may lead to local regions
containing a weakened OP, with respect to phase fluctuations. In
this connection, it is worth noting that such weakened regions
could also be nucleation centers for phase-slip phenomena, which
were seen in recent transport studies on superconducting YBCO
microstrips \cite{Jelila,Morales}.

It should also be remarked that our discussion of low-energy
Andreev states is not limited to only \emph{specular} cases. These
Andreev states could also form under \emph{non-specular}
conditions in a \emph{d}-wave superconductor, i.e. near grain
boundaries, extended non-magnetic impurities, or surface
interstitial defects
\cite{Balatsky,Tanuma,Adagideli,ZhuHu,Samokhin,Fogelstrom,PanDavis}.
Like the specular Andreev bound states, these non-specular
resonance states are essentially also based on \emph{d}-wave
Andreev interference, and should therefore also be conserved under
an applied current, as was shown in a recent theoretical
calculation\cite{ZhangIMP}. However, another calculation for this
non-specular case, done as a function of an applied magnetic
field, indicated a non-conservation of Andreev
states\cite{Samokhin2}. Further theoretical work would be
necessary to clarify this issue. Experimentally, the technique
presented in our present work, measuring the evolution of the
Andreev states under a supercurrent, may in fact provide a
definitive way to distinguish between the specular and
non-specular cases\cite{Doppler,ZhangIMP}.

Finally, we note that ZBCP were also seen on \{001\} samples,
showing the same suppression effect by currents applied
predominantly along $<$100$>$.  Figure \ref{PRL_fig3} plots an
example of such ZBCP at 4.2K. Such ZBCP have previously been
observed on \emph{c}-axis cuprate thin films \cite{MisraYazdani},
and could be explained by the presence of terrace edges on the
film surface (see inset of Fig.\ref{PRL_fig4}) which would allow
in-plane tunneling to occur on nominally \emph{c}-axis samples.
The probability for in-plane tunneling could also be enhanced by
the inherently larger matrix element for in-plane versus
\emph{c}-axis tunneling \cite{WeiPRB}.  Consistent with this
picture are instances where the ZBCP is seen together with a gap
structure at the same tip location on \{001\} samples. The main
panel in Fig.\ref{PRL_fig4} shows such a spectrum, with a ZBCP
flanked by well-developed gap edges at $\approx$$\pm$14meV. The
gap structures could be identified as the \emph{d}-wave averaged
density of states associated with \emph{c}-axis tunneling
\cite{WeiPRL,Renner,WeiPRB}. Since \emph{c}-axis tunneling does
not involve Andreev interference, the gap structures are expected
to depend only on the OP amplitude and be relatively insensitive
to the OP phase fluctuations.  What we observe in
Fig.\ref{PRL_fig4} is entirely consistent with this expectation.
Upon the application of a supercurrent, the ZBCP is suppressed as
before but without significant variation in the gap structure.
This is a remarkable observation, indicating that an applied
supercurrent could weaken the \emph{local} phase coherence while
maintaining the pairing amplitude.

\section{CONCLUSION}

In summary, using a scanning tunneling spectroscopy technique we
have observed nanoscale suppression of low-energy Andreev states by
a directly-applied supercurrent in YBa$_2$Cu$_3$O$_{7-\delta}$
thin-film strips at 4.2K. The distinct non-conservation of Andreev
states indicates that this phenomenon cannot be easily attributed to
extrinsic effects such as Joule heating, vortex flow or excess
quasiparticles. We have analyzed this suppression in terms of
order-parameter phase fluctuations, using a model based on the
\emph{d}-wave BTK formalism. Qualitative agreement between our model
and data suggests that the supercurrent may induce local phase
fluctuations in the high-temperature superconducting cuprates.
Further work is needed to elucidate this nanoscale phenomenon.

\begin{acknowledgments}
This work was supported by grants from NSERC, CFI, OIT, MMO/EMK,
and the Canadian Institute for Advanced Research in the Quantum
Materials Program.
\end{acknowledgments}


\begin{references}

\bibitem{Tsuei} C.C. Tsuei and J.R. Kirtley, Rev. Mod. Phys.
\textbf{72}, 969 (2000); C.C. Tsuei, J.R. Kirtley, G. Hammerl, J.
Mannhart, H. Raffy, and Z.Z. Li, Phys. Rev. Lett. \textbf{93},
187004 (2004).

\bibitem{VanHarlingen} D.J. Van Harlingen, Rev. Mod. Phys.
\textbf{67}, 515 (1995).

\bibitem{Hu} C.-R. Hu, Phys. Rev. Lett. \textbf{72}, 1526 (1994).

\bibitem{TanakaKashiwaya} Y. Tanaka and S. Kashiwaya, Phys. Rev.
Lett. \textbf{74}, 3451 (1995).

\bibitem{ABSrevu} For reviews, see:  S. Kashiwaya and Y. Tanaka, Rep. Prog. Phys. \textbf{63}, 1641 (2000);  L\"{o}fwander, V.S. Shumeiko and G. Wendin, Supercond. Sci. Technol. \textbf{14}, R53-R77,
(2001);  G. Deutscher, Rev. Mod. Phys. \textbf{77}, 109 (2005).

\bibitem{WeiPRL} J.Y.T. Wei, N.-C. Yeh, D.F. Garrigus, and M. Strasik, Phys. Rev. Lett.
\textbf{81}, 2542 (1998).


\bibitem{Doppler} D. Zhang, C. S. Ting, and C.-R. Hu, Phys. Rev. B \textbf{70}, 172508 (2004).


\bibitem{Khavkine} I. Khavkine, H.-Y. Kee, and K. Maki, Phys. Rev. B
\textbf{70}, 184521 (2004).

\bibitem{Zapotocky} M. Zapotocky, D. L. Maslov, P. M. Goldbart,
Phys. Rev. B \textbf{55}, 6599 (1997)

\bibitem{Kabanov} V. V. Kabanov, Phys. Rev. B, \textbf{69}, 052503
(2004).


\bibitem{Aprili} M. Aprili, E. Badica, and L.H. Greene, Phys. Rev. Lett. \textbf{83}, 4630
(1999), and references therein.

\bibitem{Krupke} R. Krupke and G. Deutscher, Phys. Rev. Lett.
\textbf{83}, 4634 (1999).

\bibitem{NgaiAPL} J. Ngai, Y.C. Tseng, P. Morales, V. Pribiag, J.Y.T. Wei, F. Chen, and D.D. Perovic,  Appl. Phys.
Lett. \textbf{84}, 1907 (2004).

\bibitem{Asano} Y. Asano, Y. Tanaka, and S. Kashiwaya, Phys. Rev.
B \textbf{69} 134501 (2004).

\bibitem{Yip} D. Xu, S.K. Yip, and J.A. Sauls, Phys. Rev. B
\textbf{51}, 16233 (1995).

\bibitem{EmeryKivelson} V. J. Emery and S. A. Kivelson, Nature \textbf{374}, 434
(1995).

\bibitem{Merchant} L. Merchant, J. Ostrick, R.P. Barber, Jr., and R.C. Dynes
 Phys. Rev. B \textbf{63}, 134508 (2001).

\bibitem{Blonder} G.E. Blonder, M. Tinkham, T. M. Klapwijk, Phys. Rev. B,
\textbf{25}, 4515, (1982).

\bibitem{ChoiCampbell} H.-Y. Choi, Y. Bang and D.K. Campbell,
Phys. Rev. B \textbf{61}, 9748 (2000).

\bibitem{FranzMillis} M. Franz and A. J. Millis, Phys. Rev. B
\textbf{58}, 14572 (1998).

\bibitem{LangDavis} K.M. Lang, V. Madhavan, J.E. Hoffman, E.W.
Hudson, H. Eisaki, S. Uchida, and J.C. Davis, Nature \textbf{415},
412 (2002).

\bibitem{Minnhagen} P. Minnhagen, Rev. Mod. Phys. \textbf{59},
1001 (1987).

\bibitem{QED3} Z. Tesanovic, O. Vafek, and M. Franz,
Phys. Rev. B \textbf{65}, 180511(R) (2002).

\bibitem{Beasley} M. R. Beasley, J. E. Mooij, and T. P. Orlando,
Phys. Rev. Lett. \textbf{42}, 1165 (1979).

\bibitem{Epstein} K. Epstein, A.M. Goldman and A.M. Kadin, Phys.
Rev. Lett. \textbf{47}, 534 (1981).

\bibitem{DemlerZhang} E. Demler, S. Sachdev, and Y. Zhang, Phys. Rev. Lett. \textbf{87}, 067202 (2001).

\bibitem{deLozanne} H. L. Edwards, A. L. Barr, J. T. Markert, and A. L. de Lozanne
Phys. Rev. Lett. \textbf{73}, 1154 (1994).

\bibitem{Affleck} J. B. Marston and I. Affleck, Phys. Rev. B \textbf{39}, 11538 (1989).

\bibitem{Varma} C. M. Varma, Phys. Rev. B \textbf{55}, 14554
(1997).

\bibitem{DDW} S. Chakravarty, R.B. Laughlin, D.K. Morr, and C. Nayak, Phys. Rev. B \textbf{63}, 094503
(2001).

\bibitem{KimWen} Y. B. Kim and X. G. Wen, Phys. Rev. B \textbf{48}, 6319 (1993).

\bibitem{Devillard} P. Devillard, R. Guyon, T. Martin, I. Safi, and B.K. Chakraverty, Phys. Rev. B \textbf{66}, 165413
(2002).

\bibitem{Jelila} F. S. Jelila, J.-P. Maneval, F.-R. Ladan, F. Chibane, A. Marie-de-Ficquelmont, L. Mechin, J.-C. Villegier,
M. Aprili,  and J. Lesueur, Phys. Rev. Lett. \textbf{81}, 1933
(1998).

\bibitem{Morales} P. Morales, M. DiCiano and J. Y.T.
Wei, Appl. Phys. Lett. \textbf{86}, 192509 (2005); Phys. Rev. Lett.
(submitted).

\bibitem{Balatsky} A. V. Balatsky, M. I. Salkola, and A. Rosengren,
Phys. Rev. B \textbf{51}, 15547 (1995).

\bibitem{Tanuma} Y. Tanuma, Y. Tanaka, M. Yamashiro, and S. Kashiwaya, Phys. Rev. B \textbf{57}, 7997 (1998).

\bibitem{Adagideli} I. Adagideli, P.M. Goldbart, A. Shnirman, and A. Yazdani, Phys. Rev. Lett. \textbf{83}, 5571 (1999).

\bibitem{ZhuHu} J. X. Zhu, T.K. Lee, C.S. Ting, and C.-R. Hu, Phys.
Rev. B \textbf{61}, 8667 (2000).

\bibitem{Samokhin} K. V. Samokhin and M. B. Walker, Phys. Rev. B
\textbf{64}, 172506 (2001).

\bibitem{Fogelstrom} M.S. Kalenkov, M. Fogelstr\"{o}m and Y.S.
Barash, Phys. Rev. B, \textbf{70}, 184505 (2004).

\bibitem{PanDavis} S.H. Pan, E.W. Hudson, K.M. Lang, H. Eisaki, S.
Uchida, and J.C. Davis, Nature (London) \textbf{403}, 746 (2000).

\bibitem{ZhangIMP} D. Zhang, C. S. Ting, and C.-R. Hu, Phys. Rev.
B \textbf{71}, 064521 (2005).

\bibitem{Samokhin2} K.V. Samokhin, Phys. Rev. B \textbf{68},
104509 (2003).

\bibitem{MisraYazdani} S. Misra, S. Oh, D.J. Hornbaker, T. DiLuccio, J.N. Eckstein, A. Yazdani, Phys. Rev. B
\textbf{66}, 100510(R) (2002).

\bibitem{Renner} I. Maggio-Aprile, Ch. Renner, A. Erb, E. Walker, and O. Fischer
Phys. Rev. Lett. \textbf{75}, 2754 (1995).

\bibitem{WeiPRB} J.Y.T. Wei, C.C. Tsuei, P.J.M. van Bentum, Q. Xiong, C.W. Chu, and M.K. Wu, Phys. Rev. B
\textbf{57}, 3650 (1998).



\end{references}
\end{document}